# X-ray microtomography of heavy microstructures: the case of Plasma-Sprayed Tungsten and Tungsten-Steel MMC

A. Zivelonghi[1*], T. Weitkamp[2,3], A. Larrue[4] and J.H. You[1]

[1] Max-Planck-Institut für Plasmaphysik, Euratom Association, Boltzmannstr. 2, 85748 Garching, Germany
[2] ESRF Grenoble, 6 rue Jules Horowitz, 38043 Grenoble, France
[3] Synchrotron Soleil, L'Orme des Merisiers, St Aubin, 91192 Gif-sur-Yvette, France
[4] Institute of Biomedical Engineering, University of Oxford, Oxford OX3 7DQ, United Kingdom

*: Corresponding author: A.Zivelonghi, e-mail: adz@ipp.mpg.de

*Graphical abstract:*

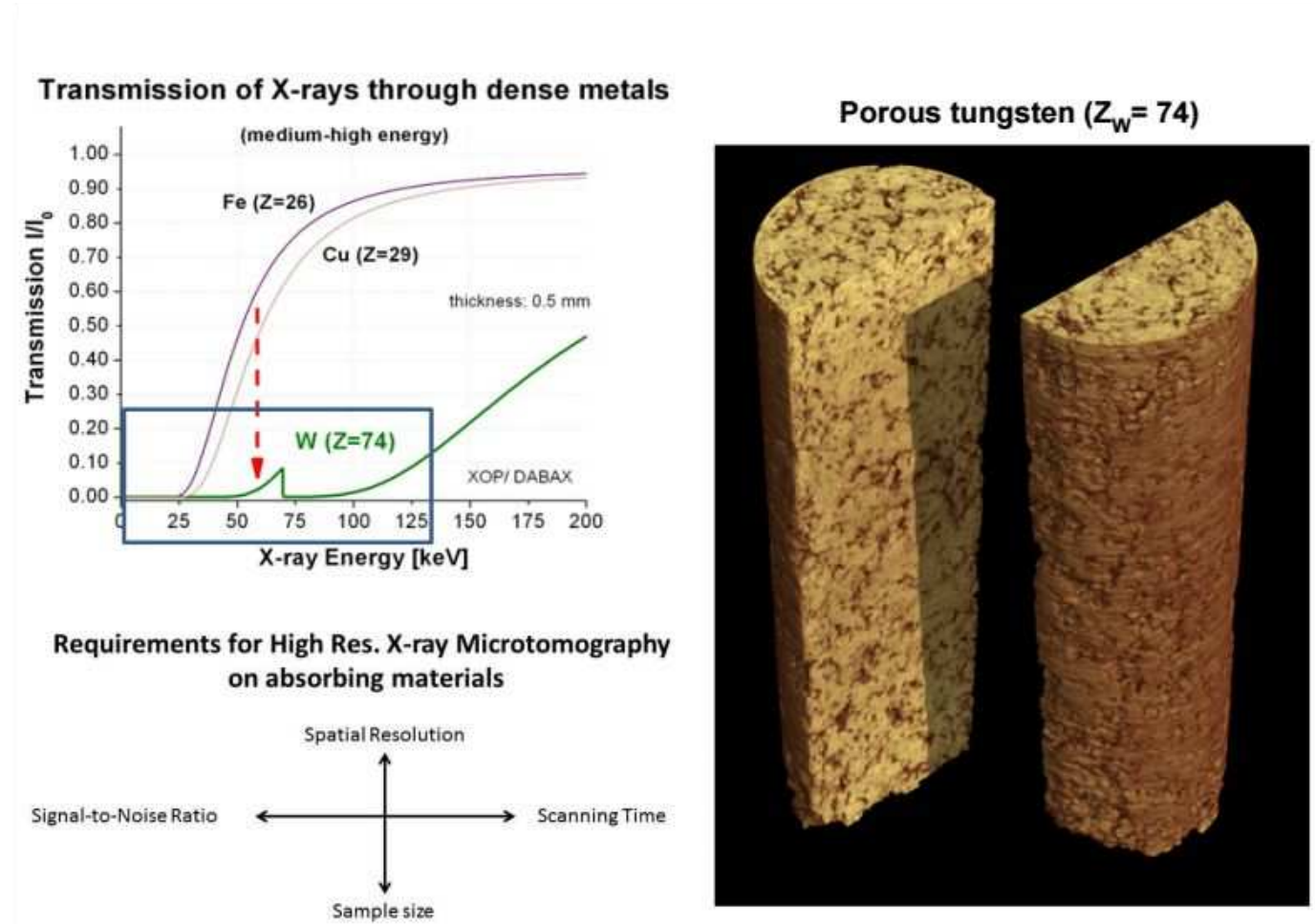

*Abstract*

In this paper synchrotron microtomography on Plasma Sprayed Tungsten (PS-W) is presented and discussed. PS-W is a challenging material for microtomography since it exhibits a random porous network at different length-scales (from nanometers to micrometers) and is hardly penetrable by X-rays. Furthermore, inner porosity causes strong internal scattering. The key challenges were, firstly, to optimize the beam parameters considering the inherent trade-off between photon energy (penetration depth) and spatial resolution and, secondly, to develop effective signal filtering algorithms. Despite the limited signal-to-noise ratio detected, large volumes of PS-W could be reconstructed with good image quality and micrometric resolution (voxel size = 1.4 μm).

As an important by-result, we report excellent image quality and higher penetration depth by applying the same setup on a ferrous microstructure, namely a 10%W/Steel MMC used as interlayer between PS-W and a ferritic/martensitic steel substrate.

The paper reports a detailed 3D morphological analysis of all inclusion types in PS-W and W/Steel, which led to disclosure of a complex connected porous network in both media. The analysis is presented in terms of multiphase volume fraction, ratio of percolation and 3D shape descriptors. 3D percolation patterns are analyzed in detail and sensitivity towards segmentation threshold for the noise-affected PS-W region is discussed. As a significant result, percolation of the porous phase in PS-W was found throughout the entire coating thickness. In W/Steel percolation was found in the perpendicular plane and interpreted as onset of delamination caused by thermomechanical stress.

# 1. Introduction

One way to generate complex microgeometries with tungsten is by means of the plasma-sprayed technique. It consists of cumulative deposition of partially molten metal droplets ejected from a plasma-gun and impacting at high speed on a water-cooled substrate target (usually steel). Plasma-sprayed tungsten (PS-W) dates back to 1961 and originates from military research [1]. Specific interest on tungsten coatings more recently developed in nuclear fusion research because of excellent thermal and mechanical performance in extreme thermal environments and because of very good compliance with plasma contamination requirements. The oxide-free vacuum plasma-sprayed tungsten (VPS-W) is now considered as candidate for armoring the first wall of future concept tokamaks [2-4].

In plasma-spraying engineering, different microstructures can be obtained by controlling plasma-spraying parameters and the substrate type. However, the 3D structure of plasma-sprayed coatings looks extremely complex with pores at different scales and a very tortuous and interconnected microtopology. In this context, *high resolution* X-ray microtomography (XMT) [5] can be successfully employed to characterize plasma-sprayed coatings and more generally thermal sprayed coatings. Because this technique is extremely sensitive to material density and to material scattering properties, previous tomographic investigation on thermal sprayed coatings were limited to elements with low or intermediate atomic number like *alumina* coatings ($Z_{Al}$=13) [6], ceramic YSZ coatings [7] ($Z_{Y}$=39, $Z_{Zr}$=40) cold sprayed titanium ($Z_{ti}$=22) [8] and cold-sprayed silver ($Z_{Ag}$=47) [9]. The present work details the first successful 3D reconstruction of plasma-sprayed *tungsten* ($Z_{w}$=74). For high Z metals, X-ray tomography is known to be hardly suitable owing to strong photon absorption in the hard X-ray regime at wavelengths substantially shorter than 0.1 nm. [10]. For example, the mass absorption coefficient of tungsten for the X-rays with a wavelength band of 0.06-0.07 nm lies between 70 and 95 cm²/g, whereas those of transition metals range from 20 to 55 cm²/g [11], which led to the X-ray transmission curves plotted in Fig. 1, where the transmission of W for accessible energies below 200 keV is from 6 to 10 times lower that of Iron ($Z_{Fe}$=26) or Copper ($Z_{Cu}$=29) at the same specimen thickness.

Strong X-ray absorption makes XMT on W very challenging, even if one considers a limited sample thickness. To overcome such limitation typical of high Z metals, the photon energy needs to be increased to several tens of keV and the first suitable energy window for W lies between 50 and 70 keV (see Fig. 1b). In porous media, however, the random microstructure introduce an additional effect: the strong internal scattering of the X-ray at the boundary between bulk and voids which combines with the high beam absorption and impose an inherent limitation to the image quality.

Moreover, for the specific case of random microstructures, the reconstructed volume should also contain a sufficient number of replicas of the random inclusions. This implies a trade-off between the applied beam energy (the higher the energy, the larger the penetrated volume) and the effective spatial resolution (which usually depletes for increasing beam energy due to depletion of detector efficiency) [12]. In addition, high photon fluence is also desired to keep the data acquisition time within reasonable limits. This originates conflicting requirements which are summarized in Fig.1c.

In this context, synchrotron radiation offers a unique possibility for tomographic characterization of difficult to penetrate media and in particular of tungsten microstructures [13, 14].

In this paper, high resolution tomography on plasma-sprayed W is presented, demonstrating feasibility of XMT on *porous* W with complex *random* microstructure. In spite of strong X-ray absorption and scattering, we report an overall good image quality, effective spatial resolution of approximately 3 μm (voxel size: 1.4 μm) and a maximum penetration depth of 0,625 mm. The measurement was performed with monochromatic synchrotron light at 52keV at the beamline ID-19 of the ESRF. Noise-affected raw images required dedicated image processing to perform automatic segmentation and quantitative analysis. This procedure (based on image splitting in the frequency domain and pass-band filters) is reported in paper.

High resolution imaging of the porous phase led to disclosure of a considerable amount of *open porosity* in PS-W. Complex percolation patterns across the entire coating thickness (>1.1mm) are revealed and their intricate geometry is analyzed in terms of 3D shape descriptors, bounding box distribution and percolation analysis. In the particular case of PS-W, percolation may strongly depend from the segmentation threshold used for the identification of the pore structures since these are affected by low SNR. This dependence was also investigated and is discussed in text.

As a secondary result, we report excellent image quality, improved contrast and higher penetration depth by applying the same optical setup (optimized for PS-W) on a less absorbing *ferrous* microstructure, namely a 10%W-Steel metal matrix composite used as interlayer between the PS-W coating and the target substrate.

Due to its importance in industrial applications, the realm of steel materials is currently extensively investigated both via standard XMT [15-17] and synchrotron XMT [18-21] . However, a tomographic investigation of W/Steel composite (more difficult to penetrate) is still lacking to the authors' knowledge. The paper reports successful microtomography on millimeter samples of ferritic/martensitic steel alloy with dispersed W particles. In its thermal history the interlayer was subjected to large temperature drops and developed a considerable amount of damage porosity at the interface with the substrate and at the boundary with the enclosed W particles. The complex three phase microstructure after thermal load is presented and analyzed.

## 2. Experiment

2.1. General remarks

The largest amount of porosity in PS-W is due to bulky globular pores reaching up to hundreds of micrometers and distant from each other tens of micrometers. For a statistically significant tomographic imaging of this material a sufficiently large size of the specimen is needed. In the case of a PS-W coating, the minimum size of the representative volume element has been estimated to be roughly (0,3 mm)³ [22].

A sufficiently large ROI is therefore an important requirement. On the other hand a plasma-sprayed coating typically exhibits a large number of fine crack at the contact interfaces between solidified droplets. The cracks usually depart from bulky pores and have thickness ranging from several micrometers to several nanometers. A sufficiently high spatial resolution is also critical to detect the dual-scale pore population. On the other hand the optically "harsh" environment (strong X-ray scattering and absorption) requires longer exposure time to improve the SNR. The technical challenge was to find a viable trade-off between two couple of conflicting requirements (Fig. 1): size of the tomographic reconstruction vs the spatial resolution on the one side and high SNR vs reasonable scanning time.

2.2 Materials

PS-W in the form of *Vacuum* plasma-sprayed tungsten (VPS-W) was provided by PLANSEE AG and produced from cumulative deposition of tungsten droplets with an $Ar/H_2$ plasma torch in a low oxygen environment (vacuum $10^{-4}$ mbar, $O_2$ <1.6 at% Fe, Ni, others <0.2 at% [23]). The partially molten droplets impinged on a water-cooled steel substrate (a ferritic/martensitic 9Cr steel called Eurofer97 with W (1%), Mn (0.6%), V (0.2%) and Ta (0.1%) as major alloying elements [24]). In order to reduce thermal stress and to improve adhesion, a thin plasma-sprayed W/Steel interlayer (0.3 mm) was inserted between substrate and coating via codeposition of tungsten and steel powder. VPS-W density (14.58 g/cm³) and porosity ($22_{\%vol} \pm 2$) were estimated by microgravimetric balance on round specimens of the isolated layer (r=0.5cm± 0.01, h=1mm± 0.05). Conical specimens of VPS-W deposited on steel were cut from large coated tiles by means of the electric discharge cutting technique (details in Appendix).

2.3 Synchrotron radiation parameters

The experimental campaign was conducted at the beamline ID19 of the European Synchrotron Radiation Facility (ESRF) in Grenoble using a high-resolution microtomography setup equipped with a multilayer monochromator [25]. A full technical description of the beam optics at the time of the experiment can be found in a dedicated publication [26]. Details of the experimental parameters applied to the present measurement are provided in Table 1. The sample diameter was ranged from 0.3 to 0.7mm along the z-axis. Since we were not interested in fast tomography, but rather in the maximization of SNR a relatively large exposure time was set (0.5-1s). For one tomogram 1500 up to 2000 projections with an exposure time of each were collected over a rotation angle of 180°. The image detector had an effective pixel size of 1.4×1.4 µm² and a square field of view whose size was 1.4 mm width and height. A europium-doped gadolinium gallium garnet (GGG:Eu, $Gd_3Ga_5O_{12}$:Eu) scintillator was used. The selected photon energy was the best empirical trade-off between penetration depth, spatial resolution and efficiency of the scintillator.

In Fig. 1b, the calculated curves of the beam transmission ratio of dense tungsten are plotted as a function of photon energy for three different penetration thicknesses (0.25, 0.5, 1 mm), respectively. This graph shows that for a thickness larger than 0.5 mm and at a photon energy below 50 keV the transmission becomes fully negligible[1]. Considering that the technically feasible minimum thickness of the porous tungsten specimen is about 0.25 mm, the penetration depth (thus the specimen thickness) was ranged from 0.25 mm to 0.5 mm. A second limitation is imposed by the GGG:Eu, $Gd_3Ga_5O_{12}$:Eu scintillator which showed a dramatic drop in efficiency (ratio of detected photons to the total number impinging upon the detector) for energies above 52 KeV (the absorption K edge of Gd is at 50.2 keV [5]). Thus the optimum between photon transmission and detection was at 52 KeV and the beam energy was fixed accordingly to this value. Full sample reconstruction was obtained by composition of two scans over the sample height.

2.4 Scanning of the W/Steel region

In order to highlight the potential on ferrous materials, a ferrous transition zone was also scanned with the same monochromatic setup at 52keV. A 6%W/ Steel metal matrix composite was prepared as an interlayer between the PS-W coating and the steel substrate. The steel matrix was made of EUROFER87, a special low-activation steel alloy developed from nuclear fusion research [2].

---

[1]. Specimen thickness lower than 0,25 mm had also lead to non representative volume of the stochastic material.

As expected, a much larger penetration depth was achieved across the W/Steel region (ca 1,3 mm vs 0,6 mm for the PS-W, although higher penetration depth in W/Steel were not tested).

## 3 Results and Discussion

### 3.1 Tomographic reconstruction

Tomographic reconstructions based on raw data of a region of PS-W are presented in Fig.2 and 5 (volume rendering tool, VGStudio Max 2.0). The image reconstruction was carried out by filtered back-projection algorithm. The characteristic porous microstructure of PS-W could be well reproduced. The complex morphological features of the microstructure consisting of randomly shaped bulky pores and microcracks are reproduced with a high precision. Very thin nanocracks whose thickness is smaller than the resolution limit of the scintillation detector (1.4μm/pixel) are not detected and thus missing in the tomographic image. However, micrometric crack-patterns are clearly visible to human eye. The internal boundaries between the dense material domain and the pores look slightly fuzzy due to the internal scattering of the beam. Another effect of the internal scattering is that the large voids appear in a relatively low gray-scale level on the tomograms rather than in black as it should be in the ideal case. For crack-like defects the gray tone increases towards the background value of dense tungsten. This causes a large overlap between the distribution of pixels belonging to crack-like defects and those belonging to dense tungsten. This geometrical noise ultimately lead to difficulties in automatic segmentation of the ultra-thin nanocracks (thickness < 1 μm) which were excluded from the segmented models.

Another factor affecting the imaging resolution is the scattering of the penetrating beam inside the medium. Strong internal scattering leads to a reduction of the signal-to-noise ratio (SNR) blurring the contrast of the porous inclusions. In the energy interval considered for synchrotron microtomography (25-100 keV) photon scattering may be very high.

A final issue is related to beam hardening. This effect is caused by the nonuniform absorption of X-rays at different photon energies. The X-rays of lower energies are absorbed more effectively in the medium than those of higher energies. Since a penetrating beam becomes progressively 'harder' as it advances through the medium, such a preferred attenuation of lower energy photons is depth-dependent and takes place more intensively in the central region of a cylindrical specimen with a circular cross section. Beam hardening is especially affecting so called white beam configurations including a large energy spectrum. In the present study we choose monochromatic light to limit this effect.

On the other hand, excellent image quality is visible on reconstructions of W/Steel (Fig. 4). Here, all three microstructural phases (W, Steel and the pores) are clearly distinguishable and reconstructed with very high contrast at voxel resolution (1,4 μm/vx). As different from the PS-W case, in the case of W/Steel we could directly perform threshold segmentation from raw data without the need for image filtering.

3.2 Image Processing

Because of low SNR, 32bit raw stacks of PS-W were processed prior to threshold segmentation in order to remove artifacts and enhance contrast. Mathematical details of image processing are described in the appendix. Border artifacts were a-priori neglected because ROIs used for calculations were always at least 10 pixel far from the real sample edge.

Fig.6a-c illustrates three different variants of a phase-segmented tomogram in PS-W before (a, b) and after signal filtering (c). Red domains indicate the empty volumes (pores or cracks) selected by thresholding on the image grayscale histogram. In (a) a lower threshold value than in (b) was applied, while in (c) the same value was applied as in (a). By comparing (a) and (b), one sees that the presence of poorly contrasted porous regions at the boundary to dense W makes not feasible the choice of an optimal threshold value from raw images causing either overestimation or underestimation of the porous region. As previously mentioned, this fact is ascribed to both beam hardening and X-ray scattering at inner boundaries between dense W and voids resulting in a limited SNR, particularly for thin void structures. As a result of filtering better segmentation of the whole porous phase with much reduced artifacts was possible (c). The thinnest microcracks (d< 2μm),

although detectable by human eye up to the voxel size, could not be included in the automatic segmentation because they showed a gray value too close to the background image noise.

3.2.1    Threshold-sensitivity

Fig. 6d shows the correlation between the applied gray scale threshold value and the estimated pore volume fraction after image segmentation. A final fine tuning of the threshold value was necessary for correct segmentation and it relies on the measured porosity of the specimens as reference data (22% ±2). This correlation was estimated by incremental variation of the gray-scale threshold value before and after filtering. Strong impact of the threshold value on the overall porosity is demonstrated together with the beneficial effect of filtering. In fact a much lower threshold sensitivity is observed when the dedicated filters are applied.

3.3  Volume fraction analysis

For the 3-dimensional image analysis the open source software ImageJ / Fiji [27] was employed. Estimated volume fractions in W/Steel and PS-W are plotted in Fig. 7. After a porosity peak in the region close to the substrate, interlayer porosity stabilizes on ca. 9% vol. while the steel content rapidly grows to an amount close to 90% vol. This value is maintained across the whole interlayer thickness. The spheroidal tungsten particles sparse inside the steel matrix reach a relative volume fraction of up to 6% vol. Then for higher z-values of the scanning region the W-phase rapidly grows towards the value in PS-W (~80%vol.). This corresponds to interruption of steel particles co-deposition and beginning of the deposition of plasma-sprayed tungsten droplets.

3.4  Percolation in PS-W

In PS-W a single very extended pore is found to percolate from the bottom part of the coating (at the interface with the W/Steel region) to the top (total height ~1mm). Fig. 8 shows the percolating pore in a subvolume of 300 μm height. A large percolating pore was found in each of the scanned samples which is a clue for extended open porosity. Previous investigation on PS-W from scanning electron microscope as reported in [22] were showing no open porosity in cross-sectional images (but obviously they could not detect percolation patterns in 3D). To quantify the extent of open porosity, the ratio of percolation (fraction of open porosity in a given direction) was calculated as $p_r = V_{i,per}/V_{i,tot}$ (volume of the phase i which is percolating over total volume of the same phase included in a volume of interest). The ratio was calculated in the OZ as well as in the OX directions (parallel and perpendicular to the coating thickness, respectively). It is remarked that all calculated ratios corresponds to a lower bound since edge pores in the region of interest were considered as separated structures [28].

As previously marked, a single very large pore with a number of tortuous branches was responsible for percolation along the OZ direction. The ratio is already very high at one fourth of the coating thickness and reaches a value of about 90% at full coating thickness. This would imply 90% open porosity along the z-axis. Along the OX direction the ratio of percolation showed a different trend (not plotted), i.e. decreasing (from 90% to 65%) for increasing x-length of the ROI (from 100 to 300 μm), but it could not be checked for larger x-values due to limited sample size. In literature, large connectivity of the pore network was also reported in ceramic plasma-sprayed coatings with lamellar microstructure (7-15% porous Yttria stabilized Zr [29]). At this stage the overall view that we get for PS-W from microtomography is a very connected globular pore network percolating along the OZ direction. However, a fully interconnected 3D network in all directions (full open porosity) may not be excluded. To prove it further investigations on larger samples at higher spatial resolution are required. It is noted that the present result is obtained neglecting the ultra-thin intersplat nano-porosity (d<1μm) since this was out of detection resolution.

The dependency of percolation ratio on the selected segmentation threshold (varied within the interval of confidence of experimentally measured porosity) was also investigated. All curves in Fig.10 indicate a major threshold sensitivity when the computed volumes are small (with a drop at 250 μm). For larger volume size (> 750 μm) the difference is actually reduced to only 5% and it becomes negligible at full specimen size.

3.5 Percolation in W/Steel

Porosity percolation was also found across the W/Steel interlayer. As different as in PS-W the largest amount of percolating porosity occurred along planes parallel to the interface with the steel substrate. A visible percolating "crack" in proximity with the steel substrate is shown Fig.9a-c and was found in all samples. The fraction of percolating porosity along the OX direction on a volume of 0,4x0,35x0,4 mm3 was estimated close to 60% (top curve of Fig. 11- note that the volume height was limited by the interlayer thickness). This result confirms the porosity peak observed at z < 50 μm by volume fractions analysis (yellow plot in Fig.7 ). A similar percolation ratio was found in the OY direction and about the same value is expected for any in-plane direction belonging to the XY plane (percolation isotropy in the XY plane). On the contrary, negligible percolation was detected along the perpendicular OZ direction (non-zero pr only for ROI length z < 200 μm as shown from the red graph of Fig. 11). This means no open porosity in W/Steel connecting the steel substrate to the PS-W layer.

Percolating cracks in the XY plane are a clear sign of ductile damage within the steel matrix caused by cycled thermomechanical stress. In fact the bilayer specimens were extracted from thicker and larger tiles which were previously loaded to several MW/m2 in a test facility which simulates nuclear fusion thermal loads [23]. The bi-layered coating (1.1mm PS-W + 0.3 mm W/Steel) experienced high thermally induced stress developed after cooling from >1000 °C to room temperature. This certainly caused high thermal stress determined by mismatch of the average thermal expansion coefficient between PS-W, W/Steel interlayer and steel substrate. As suggested from tomographic cross section (Fig.4) , the largest pores seem to originate from detachment of the hard W particles and to have coalesced thereafter. The largest pores contain or confine with at least one W particle and they are randomly located as the particles are.

3.6  3D shape descriptors

Following the method of Para-Denis and Amsellem [6, 30] the 3D morphology of all inclusions in PS-W and W/Steel was further analyzed by inertia-based shape descriptors. The descriptors $\lambda_i$ = of an object in 3D space are defined as the ratio $I_i$ /($I_x + I_y + I_z$ ) where $I_i$ (i = x,y,z) are the principal moment of inertia. Based on the inertia parameters $\lambda_i$, complex shape objects may exhibit similarity to three types of fundamental shapes: i.e spherical, flat and needle.  Fig.12a (red circles) illustrates shape descriptors distribution in PS-W as calculated in a representative cylindrical volume (height: 450 μm, diameter: 472 μm, z-axis parallel to the coating deposition direction). A very heterogeneous distribution is observed with a slight concentration in the middle-top part of the triangle towards needle type, prolate and flat structures. It is noted that the 10 largest void structures (bold marks) include 95% of the total porosity which confirms the presence of very large pores.. Higher spread of porosity shape descriptors is observed in W/Steel (Fig. 12b). W inclusions, as expected, show a more defined spherical or flat circular shape (Fig. 12c).

## 4. Summary and Conclusions

In this study we showed successful microtomography on Plasma Sprayed Tungsten and W/Steel MMC as representative of highly X-ray absorbing metal microstructures. The achievement was possible by means of high energy monochromatic synchrotron radiation and image processing. For demonstrating proof-of-principle the work was focused on three issues, namely, 1) beam energy optimization at 52 keV using a multilayer monochromator (Ru/$B_4$C) and a high efficiency GGG:Eu ($Gd_3Ga_5O_{12}$ Eu) scintillator 2) image analysis of the porous phase and removal of tomography artifacts by use of effective signal filtering 3) showing the capability of the optimized setup on a the less absorbing W/Steel microstructure.

The main achievements of the current investigation were:

1. Penetration depth up to 0.625 mm in 22% porous PS-W (average density 15 g/cm$^3$) was successfully tested. The complex microstructure of plasma-sprayed tungsten could be clearly captured by synchrotron microtomography within a reasonable measuring time (about 17min/scan) while facing high absorption and scattering conditions. 3D rendering reproduced the realistic microstructural features of PS-W and revealed a significant population of irregular and 3D interconnected pores. Detector resolution of 1.4 μm/voxel allowed human eye detection of the thinnest microcracks. 3D frequency filters on the noisy tomographs permitted automatic segmentation of the globular pore network including microcracks up to a thickness of 2.8 μm.
2. Very good tomographic quality of a mesoscopic region of a 6%W/Steel MMC (located below the PS-W layer) was achieved, with an effective spatial resolution close to voxel resolution and a penetration depth close to 1.5 mm. The observed sampling contrast for the W/Steel MMC is excellent.
3. Quantitative morpho-analysis revealed a very heterogeneous pore population and a large amount of open porosity (>90%) in PS-W percolating along the coating growing direction on tortuous 3D patterns. Percolation was also found in W/Steel but only in XY planes in proximity to the steel substrate. The percolation in W/Steel was interpreted as delamination porosity caused by cycled thermomechanical load and cooling of the coating with large temperature drop.

The results presented in this paper, combined with the most recent advances of absorption XMT on other dense media (for instance [13, 31, 32]), show a potential for microtomographic investigations of defects in dense microstructures requiring high spatial resolution and large penetration depth at the same time. They also indicate PS-W as an ideal testing material for micro and nanotomography.

**Acknowledgement**

This work was carried out within the framework of the FEMaS project supported by the European Commission. Dr. P. Tafforeau, Dr. Alexander Rack (ESRF ID19), Marco di Michiel and Mario Scheel (ESRF ID-15) are deeply acknowledged for technical support and useful discussions. The authors want to thank PLANSEE AG (Reutte, Austria) for manufactuing the coated tiles, Stangl & Co. Präzisionstechnik GmbH (Roding, Germany) for sample preparation. A.Z. express gratitude to Prof. J-H. You and Prof. Ch. Linsmeier for encouragement and support. One of the authors (T. W.) acknowledges support from the French research networks (RTRA) "Digiteo" and "Triangle de la Physique" (grants 2009-79D and 2009-034T). The measurements were carried out in beamtime allocated through the ESRF user program (proposal MA-890).

Figure 1 (Top) Attenuation behavior of X-rays penetrating in three different metals: tungsten, iron and copper (thickness is fixed to 500 μm). The transmission ratio curves were estimated based on the linear attenuation coefficient data using the XPOWER module of XOP and the Windt database in DABAX [33]. (bottom) Conflicting requirements while performing XMT on highly absorbing materials.

Figure 2 Cross-sectional view of the whole specimen reconstructed with monochromatic synchrotron light at 52 keV. The top part is plasma-sprayed tungsten (pores in dark gray, the sparse black spots are ring artifacts). The lower part is the W/Steel interlayer (W particles in gray, pores in black). The scale bar is 200 μm long. The full image height measures 1,4 mm.

Figure 3 Tomographic slice of PS-W taken at 52 keV with exposure time of 0.5s (a) and 1s (b). In the middle of (a) a large ring artifact is visible.

Figure 4 Tomographic slice of the W/Steel interlayer region Tungsten particles are shown in white, the surrounding steel matrix in gray, pores within the steel matrix in dark gray.

Figure 5 3D tomographic reconstruction of the plasma-sprayed tungsten (whole cylindrical sample). Direction is upside-down with respect to the deposition direction.

Figure 6 Three different variants of the same phase-segmented tomogram of plasma-sprayed tungsten before (a, b) and after signal filtering (c). a) and b) indicate two different phase-segmentation cases created using a high and a low gray-scale threshold values, respectively, on the unfiltered slices. In c) a clear removal of ring artifacts and pore overestimation artifacts is shown. d) Correlation between the applied threshold value and the resulting pore volume fraction after automatic segmentation.

Figure 7 (Top) Volume fractions of tungsten, steel and pores in W/Steel interlayer and part of the P-W layer. (Bottom) Cross sectional view of the corresponding region.

Figure 8 Visualization of the largest pores in a PS-W region of interest (diameter: 310 μm, height: 300 μm. ). The percolating pore is plotted in white-gray Non-white colors refer to different non-percolating pores.

Figure 9 (a) Rendering of the porous phase in W/Steel IL. (b,c) planar section in the xy plane of the 3D percolating crack at the bottom of the IL. (d,e) lateral view of the same percolating crack. Arrows indicate percolation directions (any direction in the xy plane is possible). Scale bars are 100 μm long.

Figure 10 Percolation ratio of the porous phase in PS-W along the coating deposition direction (z axis). The z values refer to increasing height of a volumetric region of interest with square basis (L=390 μm, z variable). Average on 3 different samples.

Figure 11 Percolation ratio of the porous phase in W/Steel along perpendicular directions (x,z axis). Volumetric region of interest (OX direction: y=400 μm, z=350 μm, variable x; OZ direction: x=450 μm, y=450 μm, variable z). Average on 3 different samples.

Figure 12 Distribution of shape descriptors for all inclusion types in PS-W and W/Steel (statistics on ca. 1000 items per inclusion type).

Figure 13 Sample geometry used for tomography measurement (PS-W on top, W-Steel in the middle and steel substrate on bottom).

Table 1. Experimental parameters applied to the tomographic imaging of plasma-sprayed tungsten at the ESRF beamline ID19 [26].

| **X-ray beam spectral settings and geometry** | |
| --- | --- |
| Photon energy E | 52 keV |
| Energy bandwidth $\Delta E/E$ | approx. $10^{-2}$ |
| Monochromator | Multilayer Ru/B4C, bilayer period d = 4 nm |
| Absorption filters used | Al 2.5 mm, Cu 0.25 mm |
| Source size (FWHM) | 120 μm × 30 μm (horizontal × vertical) |
| Distance from source to sample | 145 m |
| Distance from sample to detector | 100 mm |
| **Detector** | |
| Scintillator | Gadolinium gallium garnet (GGG) doped with europium ($Gd_3Ga_5O_{12}$:Eu), thickness of doped layer 10 μm. |
| Conversion optics | Objective 10×/NA 0.3; eyepiece 2×; overall magnific. ×20 |
| Electronic sensor | ESRF FReLoN 2K14 CCD, 2048×2048 pixels of 14 μm size, dynamic range approx. 11000:1 (13.5 bit) |
| Effective pixel sizeize | 1.4 μm (using 2×2 binning; effectively 1024×1024 pixels) |
| **Sample** | |
| Sample diameter | plasma-sprayed tungsten: 0.25 mm - 0.5 mm |
| | tungsten-steel composite: 0.5 mm - 1.5 mm |
| **Tomography scan parameters** | |
| Angular range scann | 180° |
| Number of projection ang | 2000 |
| Exposure time per an | between 0.5 and 1.5 s |
| Total time per sc | between 17 and 38 minutes |

# APPENDIX

## A1. Image processing of PS-W

A dedicated signal processing algorithm was developed to enhance the contrast of the structures of interest while preserving pore connectivity. The approach followed in this study aims at preparing the raw image data in such a way that a suitable unique gray-scale threshold value can be set for automatic phase segmentation on each tomographic slice. Automatic phase segmentation is required because of large data sets to be processed in 3D morpho-analysis.

The filtering process is based on three steps: 1. gaussian high-pass filter applied to raw tomographs whose aim is to separate the high and low frequency structures, 2. separate processing of the high and low frequency images and 3. image reconstruction. The low frequency image is smooth and discloses the bulky pores while the high frequency one reveals the thin cracks and the small noisy structures, such as ring artefacts (RA). These are removed by using a standard median filter on the image transformed in polar coordinates. A more complex image processing is then applied to the high-frequency image. It aims at increasing the contrast of thin linear (in 2D) or planar (in 3D) structures by first computing on each point a parameter related to the local planarity. This approach is described in detail in [34, 35]. Here it has been adapted to the PS-W case. The grey level values of the high-frequency image are filtered using the following formula:

$$g(x) = \frac{1}{\mu} \cdot \sum_{y \in V(x)} f(y) \cdot e^{-\frac{\|f(x)-f(y)\|^2}{t^2}} \cdot e^{-\frac{\|S(x)-S(y)\|^2}{\tau^2}} \qquad (3)$$

where $g(x)$ is the grey level of the filtered image at the point $\mathbf{x}$, $V(x)$ is a 3D cubic neighborhood around $\mathbf{x}$, $f(x)$ and $f(y)$ are the grey level values of the original image at the point $\mathbf{x}$ and $\mathbf{y}$, $S(x)$ and $S(y)$ are the values of the planarity parameter at the point $\mathbf{x}$ and $\mathbf{y}$, $t$ and $\tau$ are parameters, and $\mu$ is a normalizing term having the following expression:

$$\mu = \sum_{y \in V(x)} e^{-\frac{\|f(x)-f(y)\|^2}{t^2}} \cdot e^{-\frac{\|S(x)-S(y)\|^2}{\tau^2}} \qquad (4)$$

This approach effectively selects voxels of the same planarity, i.e. likely to belong to the same thin structure, to average their grey levels. Within the areas where no planar or linear structure is detected, the selectivity decreases and the filter is equivalent to a generic smoothing by averaging. This way, the noise is reduced in the homogenous areas and the connectivity of thin structures is preserved. Finally, the low-frequency image and the enhanced high-frequency are combined.

## A2. Sample preparation via electric discharge cutting

Conical specimens of VPS-W deposited on Eurofer97 steel by PLANSEE AG (Reutte, Austria) were cut from large coated tiles by means of electric discharge cutting (Stangl & Co. Präzisionstechnik GmbH - Roding, Germany). Details of specimen geometry is shown in Fig. A2. Because of tungsten brittleness the minimum possible specimen thickness during machining was 0.25 mm. Easy to handle specimens with a larger cilindric base (diameter 1 cm) with on top the very thin and fragile PS-W needle were prepared. Electric discharge cutting allowed to prepare samples with diameter of PS-W rangin from 0.25 to 0.65 mm (along the coating deposition direction, i.e. perpendicular to the synchrotron scanning beam). The total height of the conical needle was 1.8 mm. The cutting angle along the W/Steel region could vary considerably from specimen to specimen, originating a cross-sectional diameter ranging from 0,65 mm to 1,35 mm.

Errore. L'origine riferimento non è stata trovata.

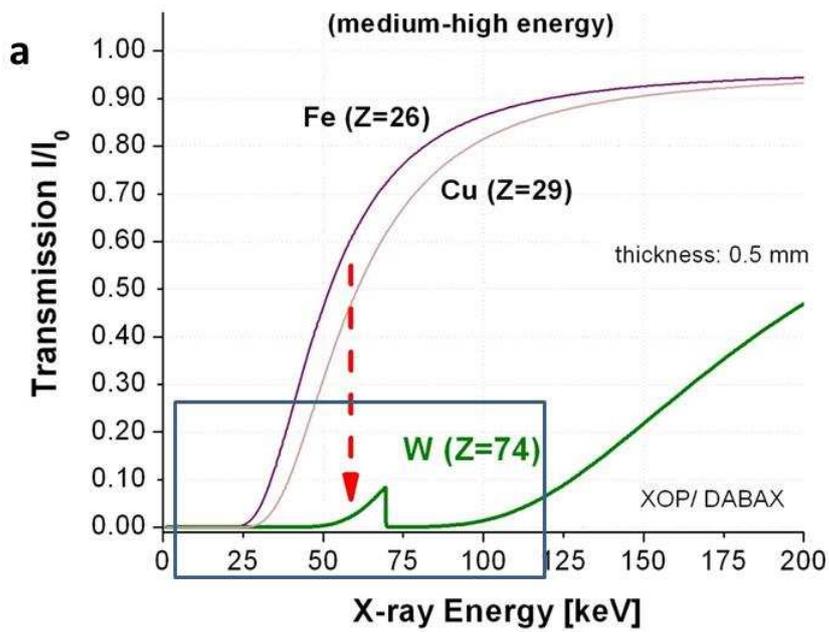

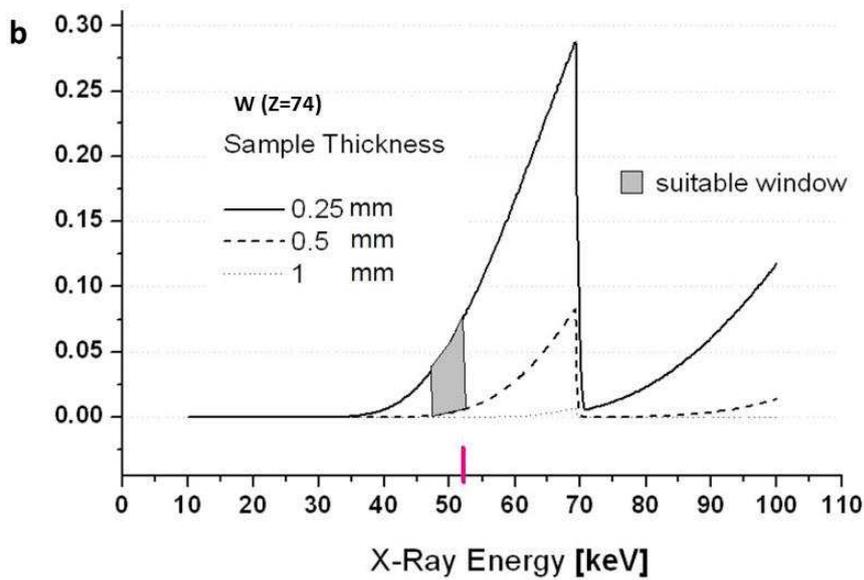

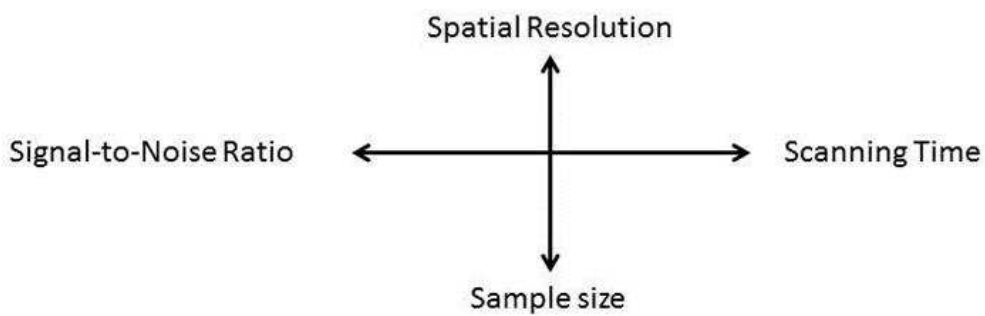

**Figure 2**

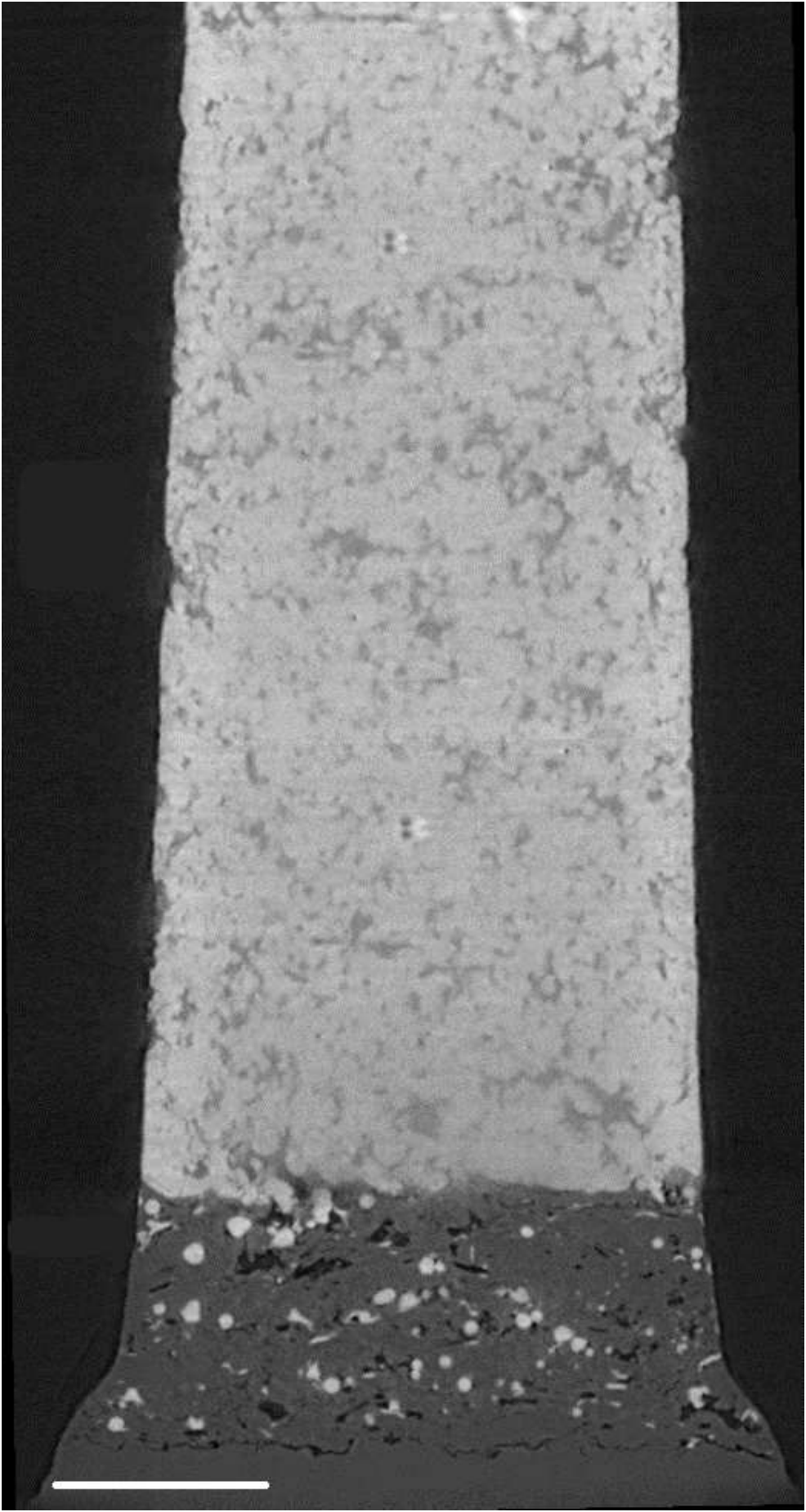

**Figure 3**

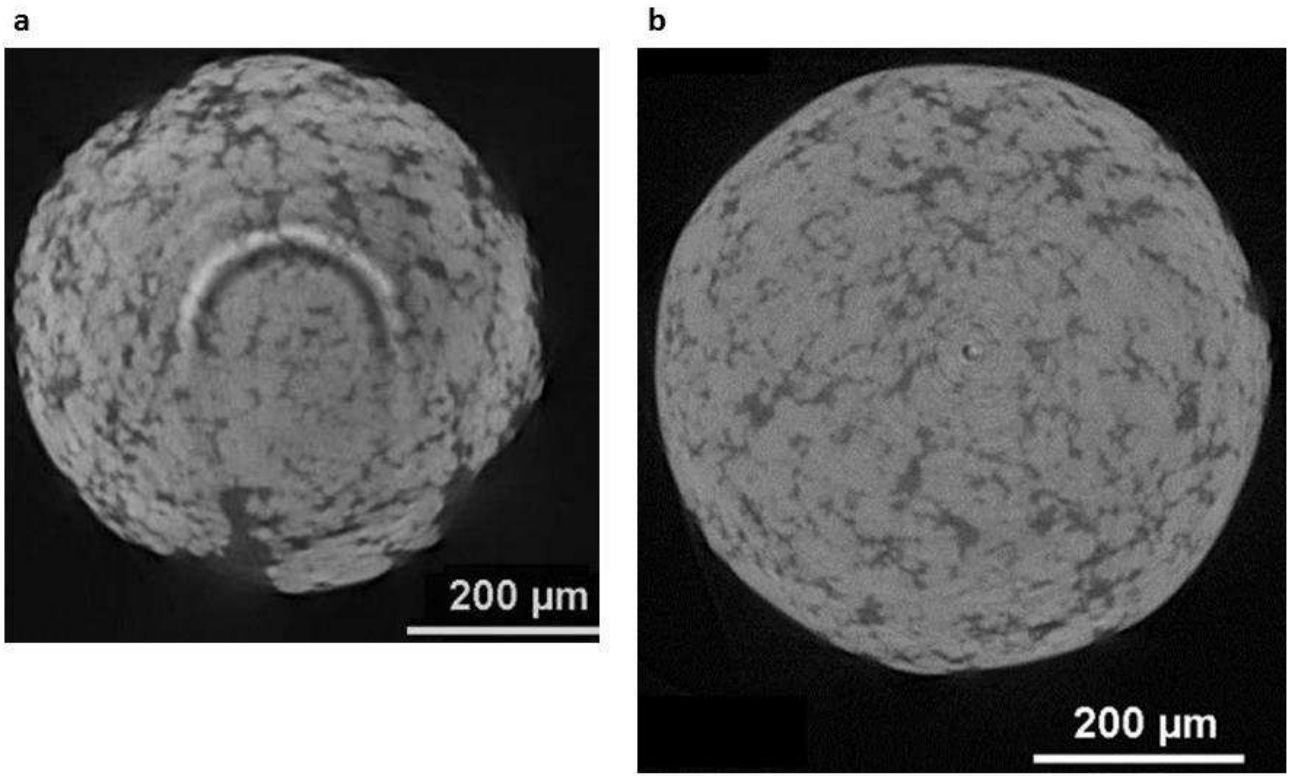

**Figure 4**

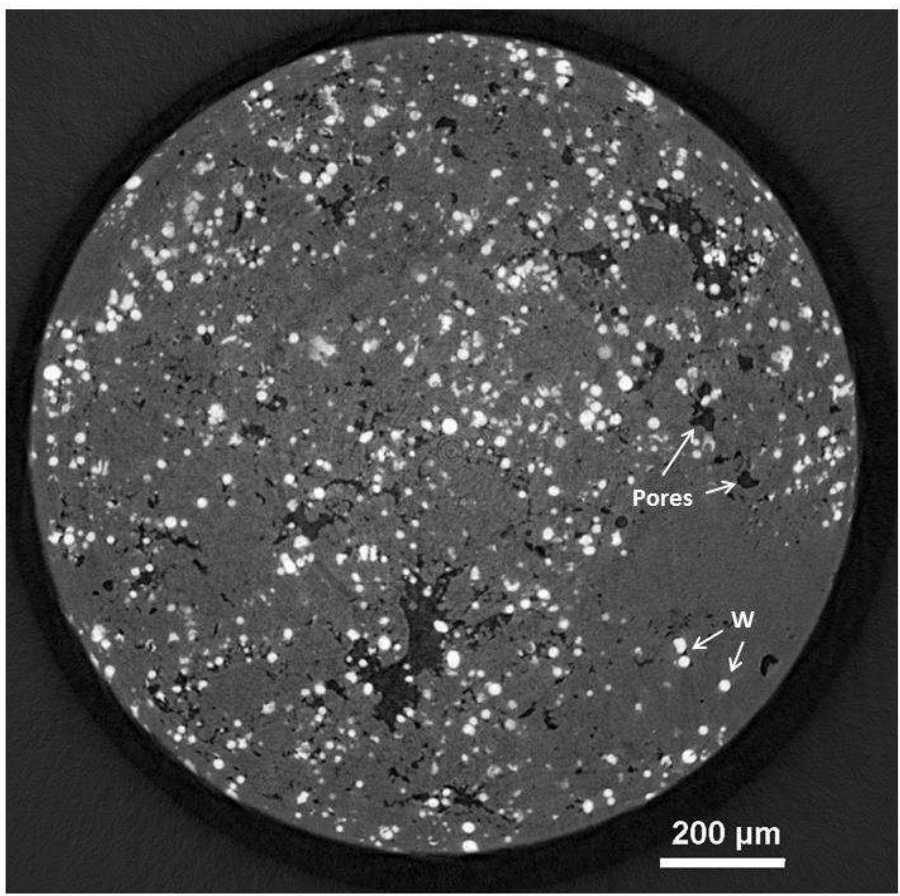

**Figure 5**

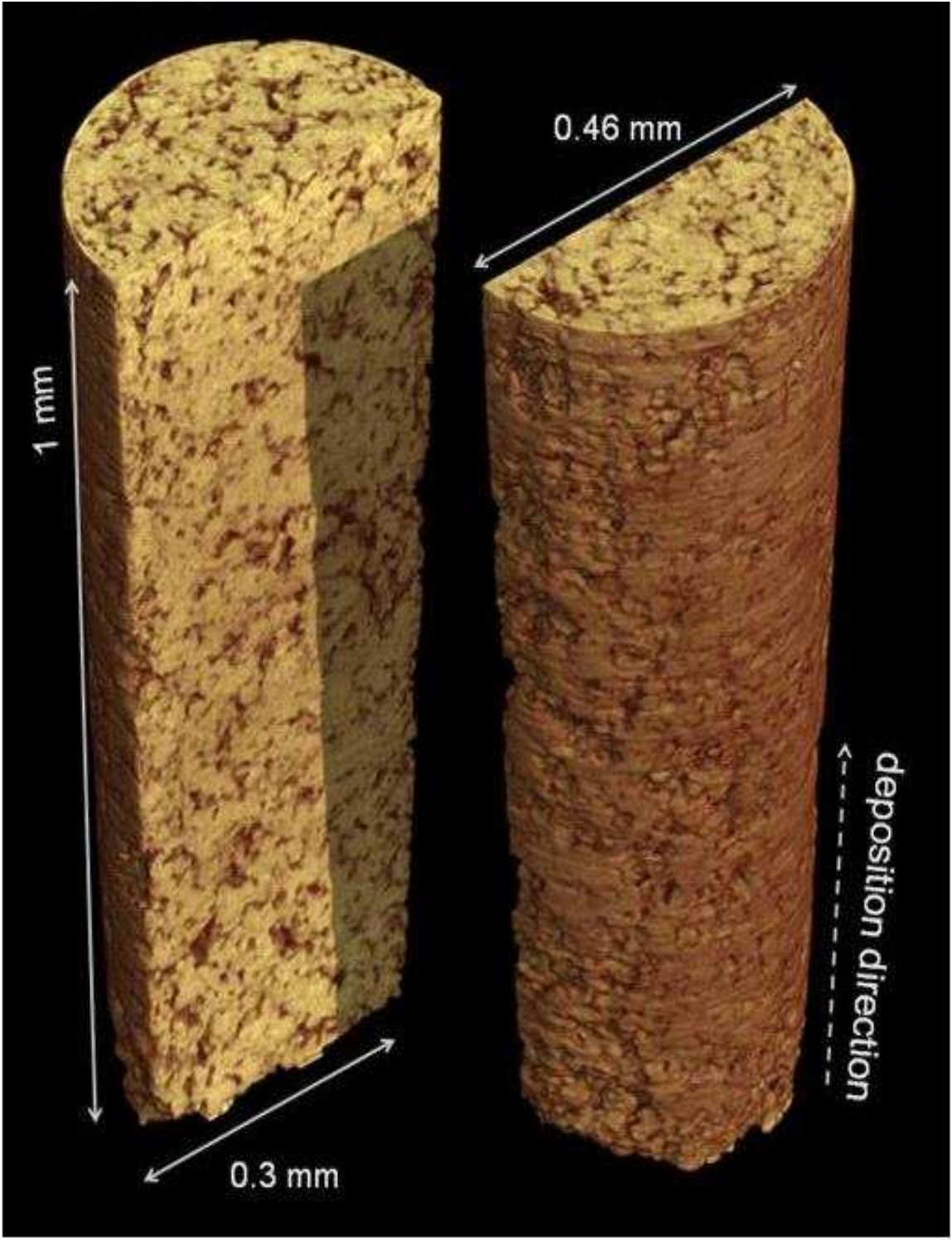

**Figure 6**

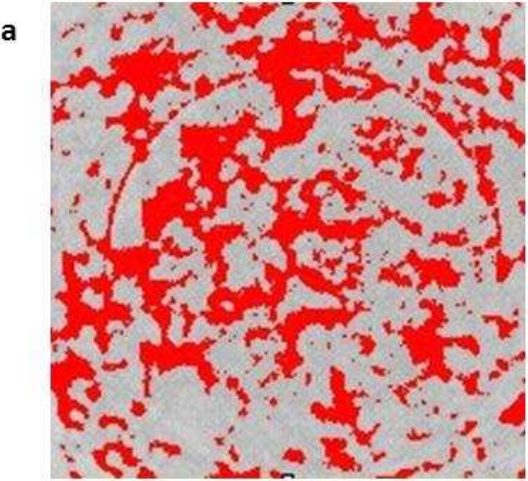 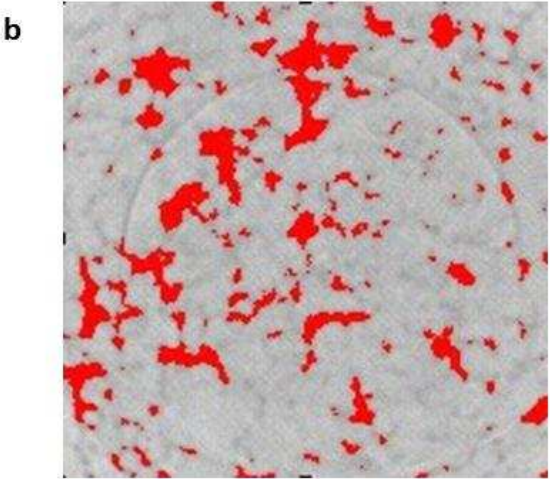
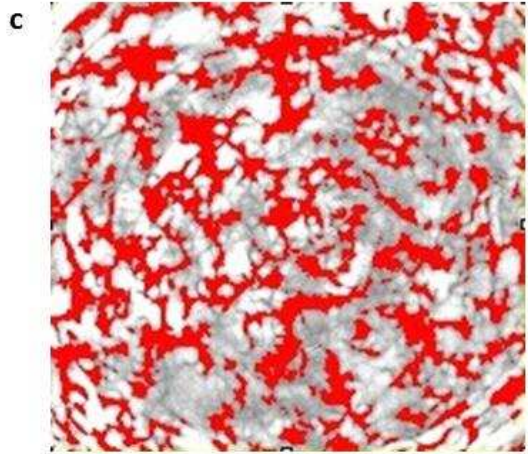 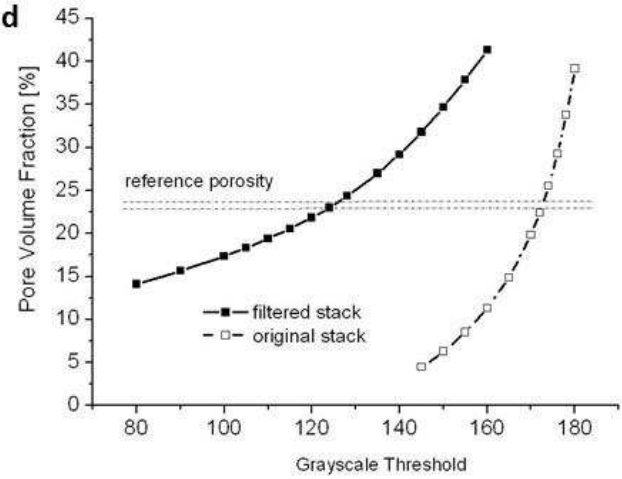

**Figure 7**

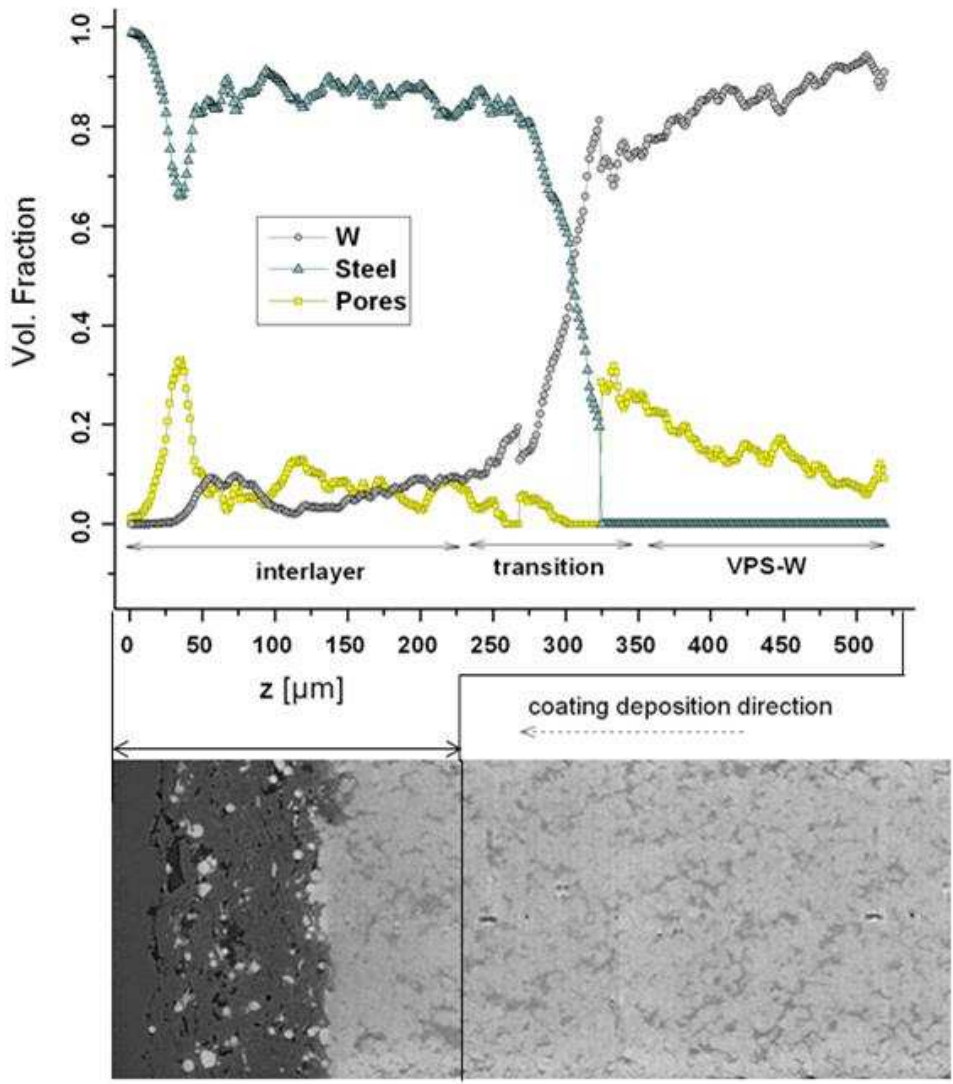

**Figure 8**

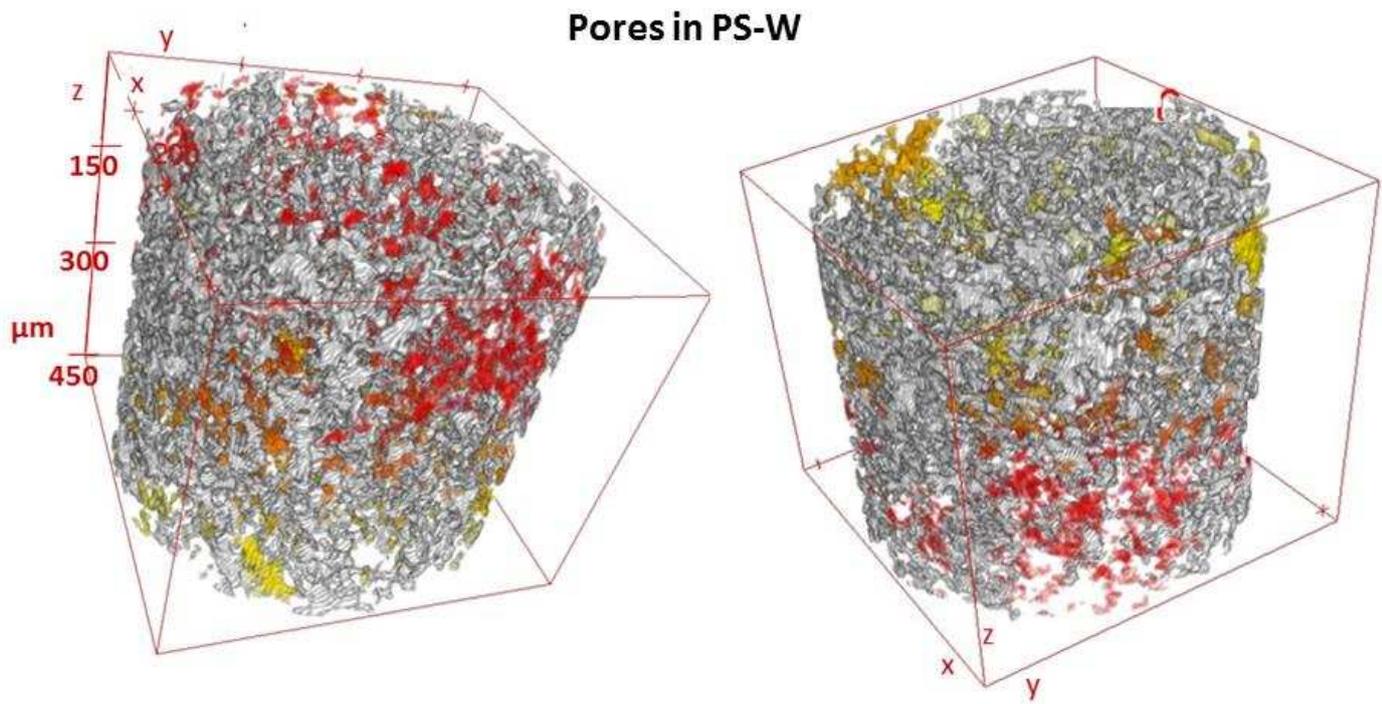

**Figure 9**

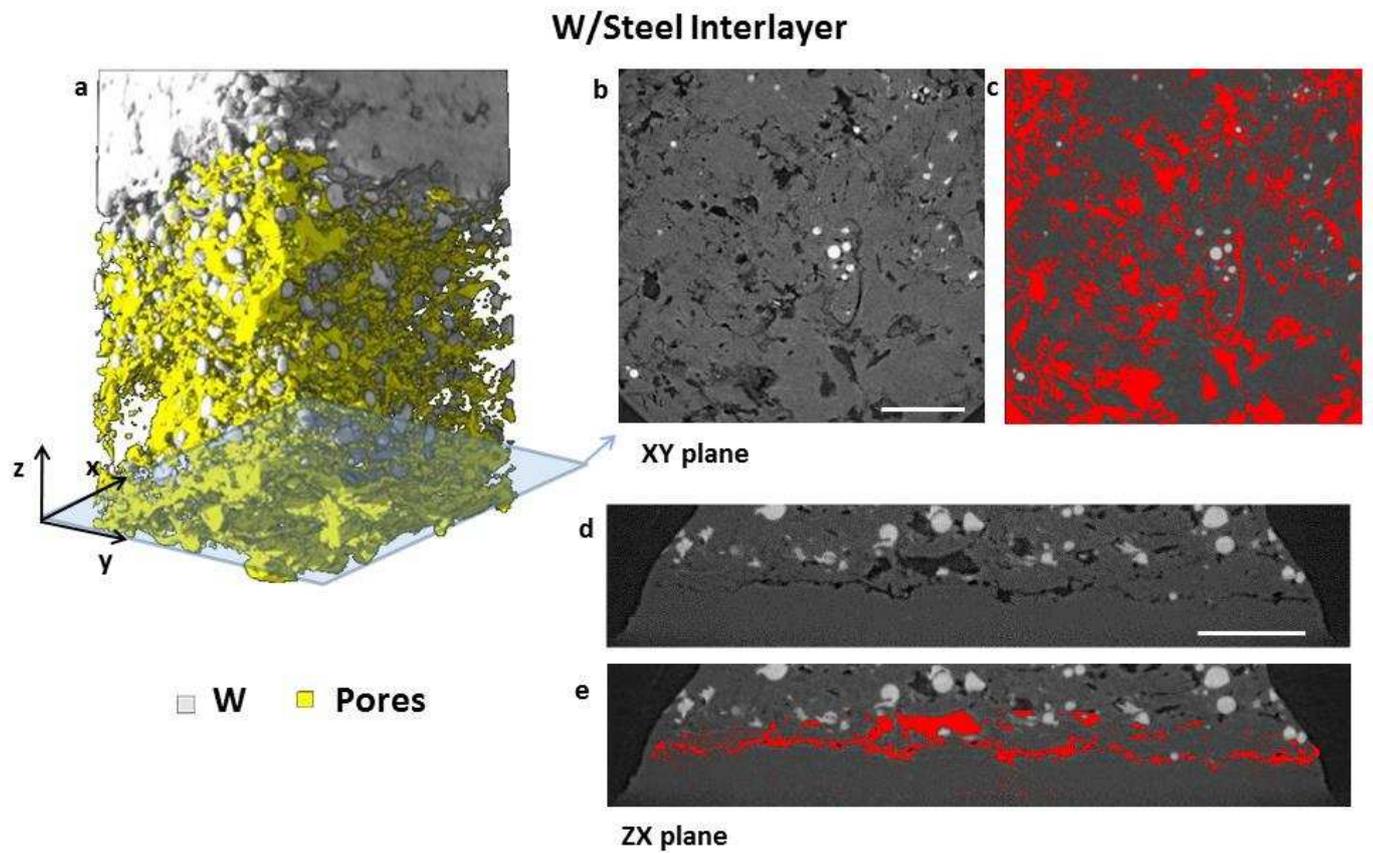

**Figure 10**

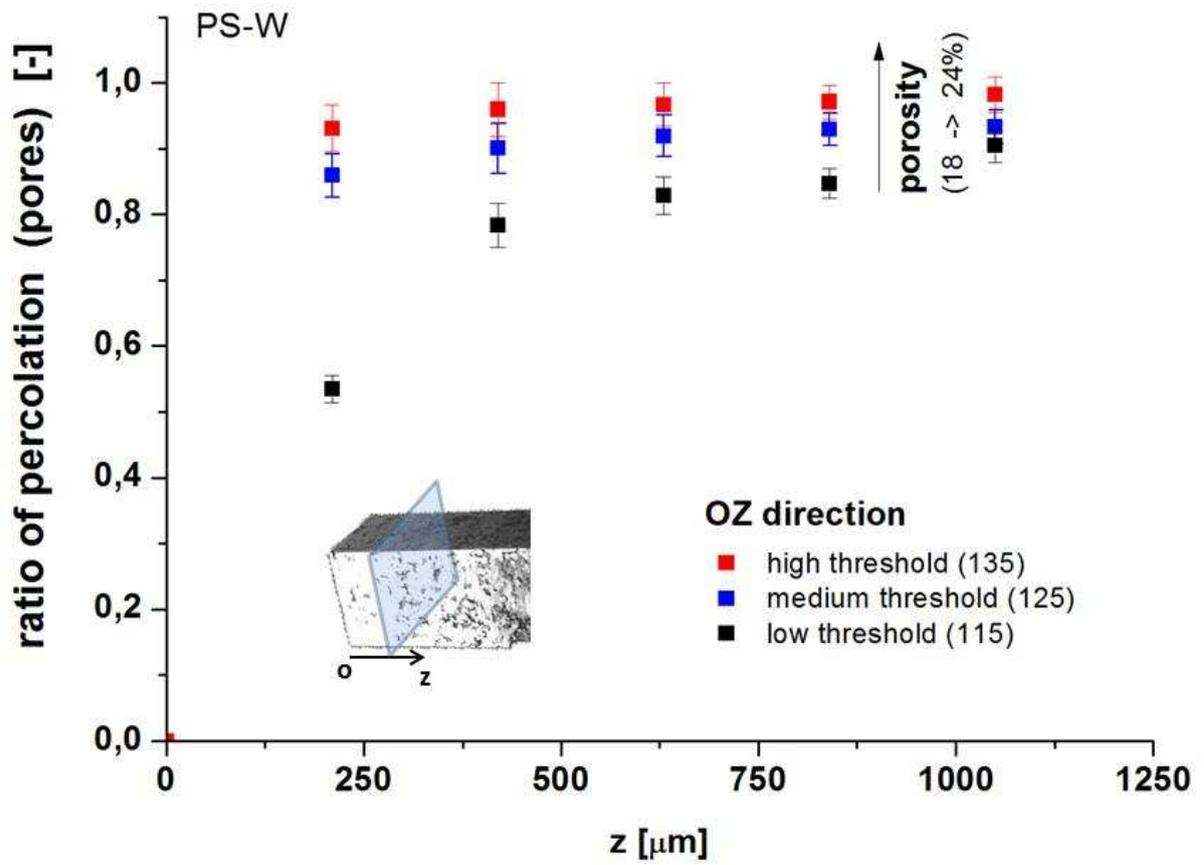

**Figure 11**

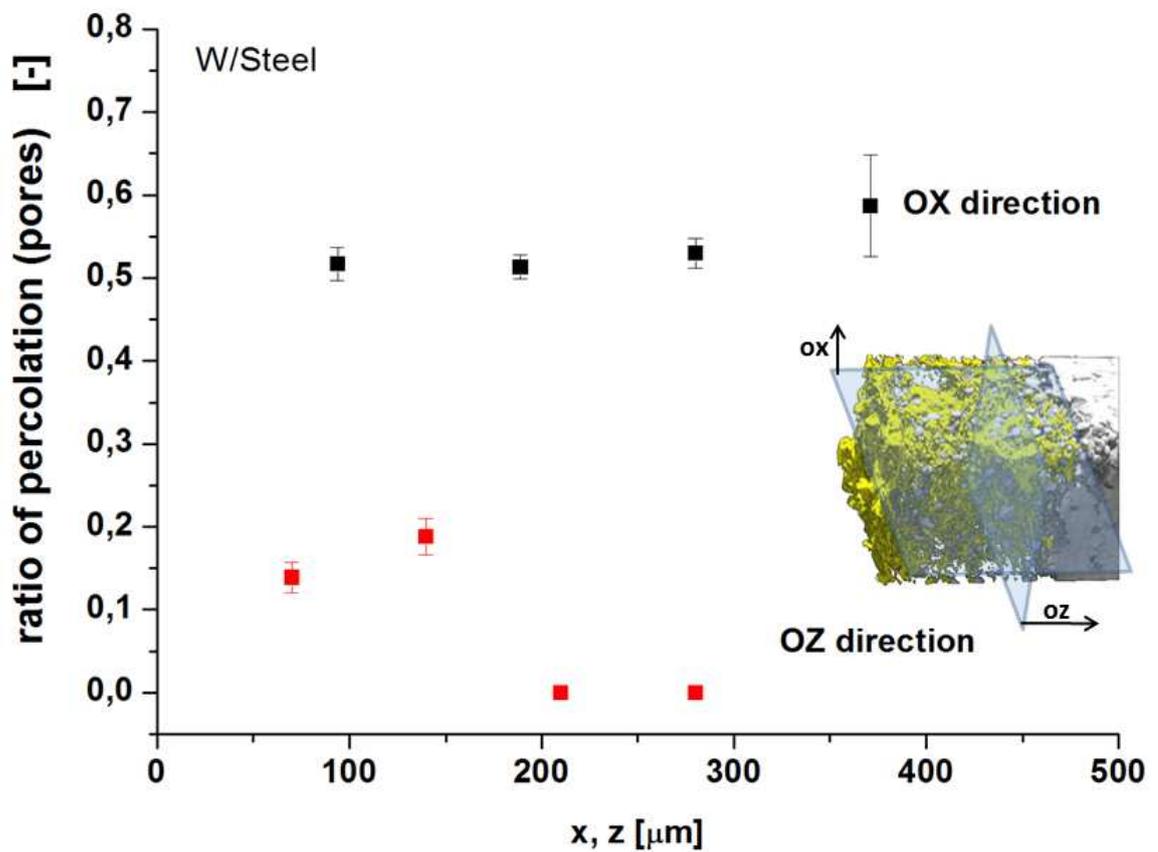

**Figure 12**

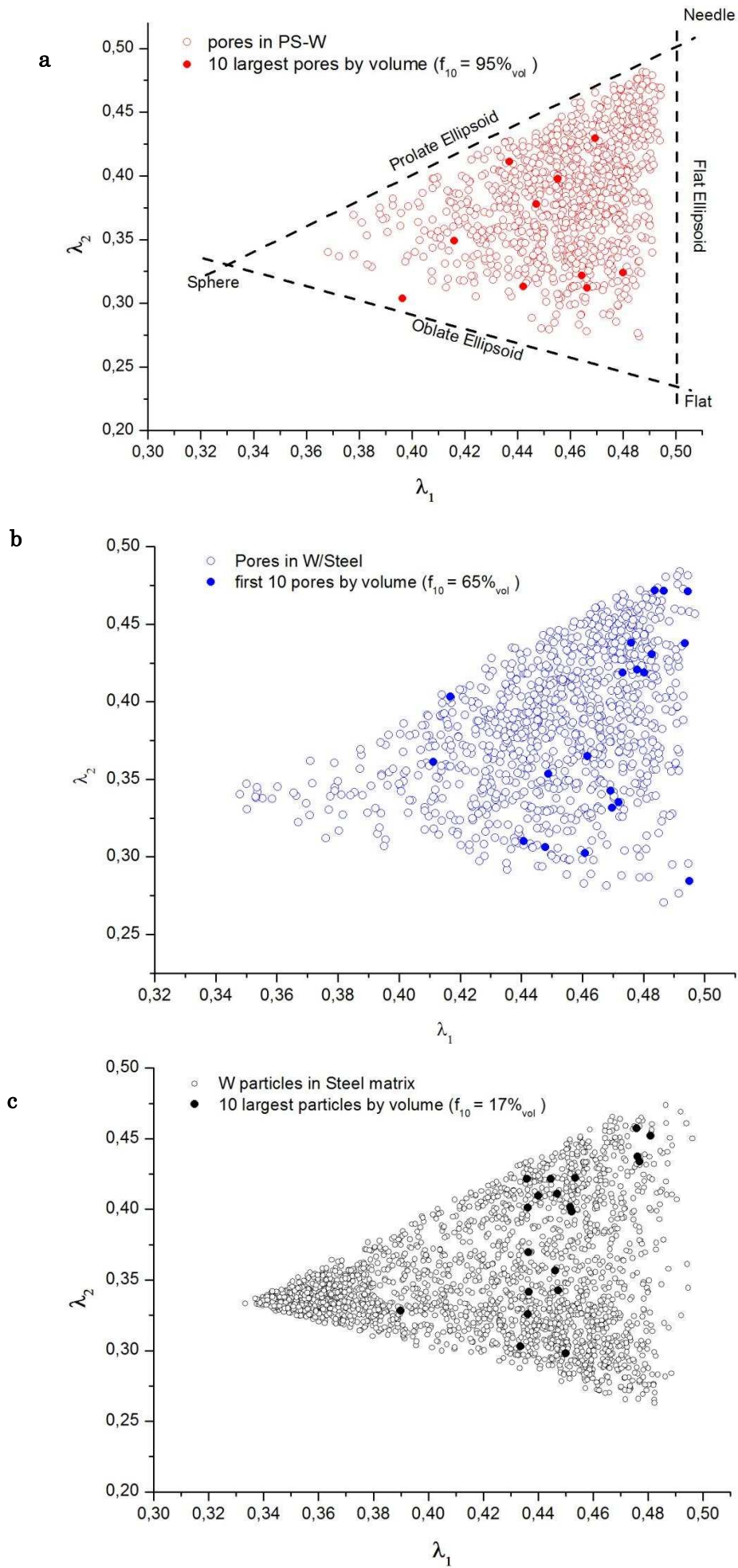

**Figure A2**

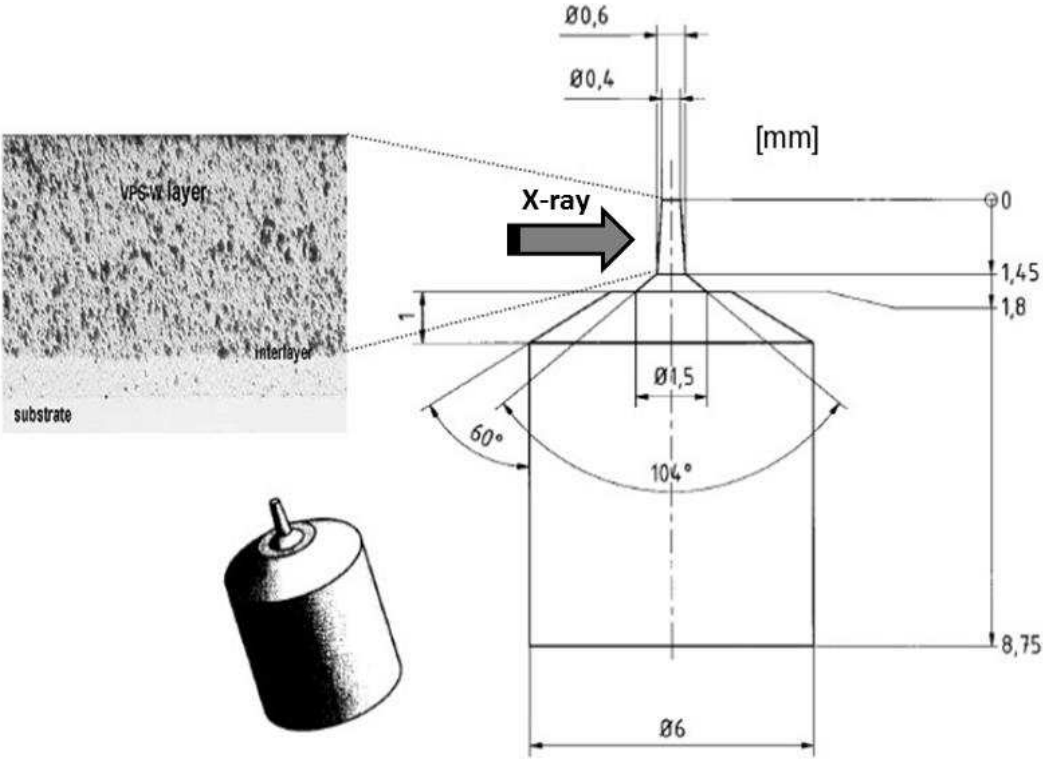